\documentclass[prodmode,acmtap]{ci}

\pdfoutput=1

\acmVolume{2}
\acmNumber{3}
\acmArticle{1}
\articleSeq{1}
\acmYear{2010}
\acmMonth{5}

\usepackage[ruled]{algorithm2e}
\SetAlFnt{\algofont}
\SetAlCapFnt{\algofont}
\SetAlCapNameFnt{\algofont}
\SetAlCapHSkip{0pt}
\IncMargin{-\parindent}
\renewcommand{\algorithmcfname}{ALGORITHM}
\usepackage{csquotes}

\title{Simulating Self-Organization during Strategic Change: Implications for Organizational Design}
\author{ANANYA SHETH and JOSEPH V. SINFIELD \affil{Purdue University}
}
\begin{abstract}
Intentionally left blank.
\end{abstract}

\begin{document}

\maketitle

% Head 1
\section{Introduction}

Self-organization- a characteristic of complex adaptive systems (CAS) has been explored in organizational research- in management theory \cite{mihm2003problem,Foerster1984}, firm internationalization \cite{chandra2017firm}, organizational design \cite{clement2017searching}, and strategic change \cite{foster2015network}. Newer organizational forms such as networks and zero-hierarchy companies that hold the promise of self-organization are gaining prominence \cite{puranam2014s}, and theoretical organizational modeling is a useful technique to study them proactively via simulation \cite{puranam2015modelling,simon1976substantive}. In this paper, we introduce a nature-inspired model to understand self-organization of collaborative groups in three archetypal organizational designs- i. fully-networked, ii. siloed, and iii. dynamic, where each design controls intra-managerial communication in specific ways, and each member has reactive or perceptive behavioral tendencies.

% Head 2
\subsection{Why a nature-inspired model for human CAS?}

We are perennially in search of improved ways of organizing such that CAS components act coherently without explicit communication \cite{simon1996sciences}. Nature reveals numerous examples: ants- collectively establish ideal pathways to food, bees- work together to hive, fish- school to protect themselves from predators, and human immune system cells- multiply at intrusion for immunity. These tasks require complex coordination but are achieved with minimal communication. Furthermore, goal-oriented agents in these systems learn by cooperative mechanisms rather than competition. Factually, cooperative management has been a dominant feature of pre-industrialization human organization which is a disproportionately large part of our existence \cite{VonRueden2015a}. Even in today's large-scale enterprises, often termed \enquote*{mechanistic}, there exist goal-oriented collaborative groups. The essence of self-organization is response to stimulus with minimal inter-agent communication, and because organizational design governs communication channels, in this paper, we hypothesize that comparing cooperative learning patterns of individually biased actors operating in differently structured organizations \cite{granovetter1985economic} via simulation can provide insight into organizational design.

\subsection{A multilevel system}

Agent-based models (ABMs), which are useful in simulating complex social systems \cite{miller2009complex,schelling1971dynamic}, take a bottom-up approach i.e., rules governing agent behavior are defined (bottom) that in aggregation generate system behavior (top). Humans- the agents of our CAS are boundedly \cite{simon1991bounded} as well as ecologically rational \cite{Todd2012} i.e. they make decisions according to what they know and according to the environment in which they are acting. We simulate conditions of change in enterprise strategy such that a new strategic vision is set at the top that is then to be explored and implemented by members of a cooperative group at the lower levels. Tversky and Kahneman \citeyear{Tversky1974} have highlighted individual level biases in humans that affect the action decisions to be made by group members. We focus on three such biases- i. Personal experience i.e., the members' tendency to lean on their own previous learning, ii. Prestige-bias i.e., the members' tendency to follow other more successful members of the cooperative group, and iii. Inertia i.e., the members' tendency to remain at status-quo. Cooperative groups often have flat structures without formal hierarchy between group members as would normally exist (formally represented in organizational charts) between members of a standard team. Rather, in such groups, individual action decisions are based on performance outcomes of the individual and that of the group. The individual and group performance is defined by how ‘far’ the individual and the group is from the strategic vision. The system is temporal and as it progresses, members' search for the mandated goal and adaptively learn paths to it. Hence, the model measure is the number of iterations it takes individuals as well as the groups to achieve the common goal. Note that this is guided search and not a random search i.e., at each recursion a criterion helps individuals evaluate their solutions. Hence, overtime, the groups' outcome tends to move in the direction of the goal due to accumulation of individual and group memory. Further, a member acts on the received individual performance feedback in two hypothesized manners- 1). Reactively- i.e., by immediately shifting behavioral tendencies with performance feedback, and 2). Perceptively- i.e., by gradually responding to accumulated feedback, yet moving to reactive behavior as (and if) performance pressure increases with time. In addition, the three organizational designs constrain members' communication. For instance, all members of a fully-networked group such as builder's of an open-source digital library may communicate freely whereas those in a siloed structure such as a traditional industrial conglomerate might not. Similarly dynamic organizations having regular reshuffling of teams might lead to novel scenarios. Therefore, self-organization in the three designs can be compared.

\section{Model}

We choose a model mechanism that is inspired from the kinematic motion of bird flocks. Hence, there are spatial and motion elements as shown in the position and velocity equations 1 and 2 respectively. The model search space is a 25-dimension binary space and each strategy is a 'position' in that space represented by a string vector of binary bits. Each bit is abstracted as a strategic dimension such as \enquote*{target some market}, \enquote*{produce economically}, \enquote*{undercut market price}, \enquote*{maintain wallet share}. The fitness of each member is how far s/he is from the ideal strategy. This is measured by the Hamming distance \cite{warren2013hacker,steane1996error} and stored in the individual's memory. Memory affects the individual's decision in the following three ways:

\subsection{Self-belief}
Organizationally, this factor captures the effect of the member's own best achieved strategic positions thus far on the next strategy selection decision. In the equation of motion, this term accelerates the member in a direction. Mathematically, it is achieved by multiplying a self-belief coefficient of the agent with the difference of the agent's current strategic position and its previously achieved best strategic position. Members with reactive tendencies will readily gain/lose self-belief with positive/negative performance feedback.

\subsection{Prestige-bias}
Prestige bias is defined as the individual's bias in identifying best performing members and following them over pursuing other paths such as the individual's own. This principle was developed from the Boyd and Richerson \citeyear{boyd1988culture} theory of cultural evolution and has been known to exist in groups where learning occurs. Similar to self-belief, it is an acceleration term, and is mathematically achieved by multiplying a prestige-bias coefficient of the member with the difference of the agent's current strategic position and the hitherto globally achieved best strategic position. A member having received negative performance feedback is likely to follow a more successful (prestigious) member.

\subsection{Inertia}
This factor incorporates the members' momentum accounting for the recent most decision. The momentum can be unpacked as the member's inertia multiplied by his/her immediate previous velocity. Hence, between two recursive steps, as the member changes his/her velocity, the new velocity is affected in part by its immediate previous velocity. In the equation of motion, this is the initial velocity term. Organizationally, this is the difficulty in or obligation to not deviate radically from the \enquote*{legacy} of past decisions.

\subsection{Position and velocity update rules for members' strategic exploration}
We use kinematic logic to govern the spatial drift of the agents' strategic positions. At each iteration, equations \ref{Position update} and \ref{Velocity update} govern the model. Equations \ref{v=u+at} and \ref{org_vel} are representations of equation \ref{Velocity update} for explanatory purpose. 
\begin{equation}\label{Position update}
P_{i}(t+1)\:=\:P_{i}(t)\:+\:V_{i}(t+1)
\end{equation}
where $P_{i}$ is the strategic positioning of an agent in consecutive evolutionary states and $V_{i}$ is the agent's velocity governed by: 
\begin{equation}\label{Velocity update}
V_{i}(t+1)\:=\:W*V_{i}(t)\:+\:C_{1}[(P_{b}(t)\:-\:P_{i}(t))]\:+\:C_{2}[(G_{b}(t)\:-\:P_{i}(t))]
\end{equation}
where W is the agent's inertia, $C_{1}$ and $C_{2}$ are acceleration constants representing self-belief and prestige bias respectively,  $P_{b}$ is the agent's personal previously learned best position, and $G_{b}$ is the current position of the most fit agent, $P_{i}$ is the agent's position at iteration (t). The change in strategic positions is by means of a spatial drift and therefore, is governed by the equation of motion:
\begin{equation}\label{v=u+at}
New Velocity = Initial\:Velocity + Acceleration*Time
\end{equation}
The drift governing coefficients depend on individual member tendencies of inertia to change, self-belief, and prestige bias. Therefore, equation (\ref{v=u+at}) is re-written as 
\begin{equation}\label{org_vel}
New Velocity = Inertia*a + Self\:belief*b+ Prestige\:bias*c
\end{equation}
where \textit{a = previous velocity, b = current distance from previous learned personal best, and \newline c = current distance from most fit agent}.
\newline
% \newline
Hence, for our model, the individual's velocity is dependent on his/her immediately previous velocity and the two accelerations based on the complementary biases held by the individual. Organizationally, it signifies every member's decision of balance between willingness to change, individuality, and followership. Members in each design are uniquely biased to simulate diverse organizations.

\section{Discussion}
Simulation results reveal that rate of self-learning across organizational designs is different. A fully-networked organization with members having reactive tendencies is the quickest to self-organize followed by dynamic and siloed designs. Organizational structures with perceptive members have slower learning. However, among structures with such members, fully-networked structure take less than half the time to self-organize as compared to siloed or dynamic structures. Dynamic reshuffling of reactive and perceptive members results in different outcomes- reshuffling perceptive members does not lead to improved self-organization hinting at potential inefficiencies such as over-networking perceptive members. The work indicates that learning in goal-oriented groups varies by member tendency and communication structure, and therefore organizational design. Hence, coupling member tendencies and organizational design is a significant research focus to understanding self-organizing groups. This modeling technique can be employed to test specific hypotheses in this regard.

% Bibliography
\bibliographystyle{ciformat}
\bibliography{cisamplebibfile}

\end{document}